\documentclass[%
 reprint,
 amsmath,amssymb,
 aps,
]{revtex4-2}

\usepackage{graphicx}
\usepackage{dcolumn}
\usepackage{bm}

\usepackage[mathscr]{euscript}
\usepackage{color}
\usepackage{xcolor}
\usepackage{xspace}
\usepackage{amsmath}
\DeclareMathOperator{\arctantwo}{arctan2}

\newcommand{\degC}{\ensuremath{{}^{\circ}\mathrm{C}}\xspace}
\newcommand{\uT}{\ensuremath{\,\mu\mathrm{T}}\xspace}

\newcommand{\Hertz}{\ensuremath{\,\mathrm{Hz}\xspace}}
\renewcommand{\sec}{\ensuremath{\,\mathrm{s}\xspace}}
\DeclareMathAlphabet{\mymathbb}{U}{BOONDOX-ds}{m}{n}

\renewcommand{\deg}{\ensuremath{{}^{\circ}}\xspace}

\def\Cth{\ensuremath{{}^{13}\mathrm{C}}\xspace}

\def\Hone{\ensuremath{{}^{1}\mathrm{H}}\xspace}

\newcommand\ket[1]{|#1\rangle}
\newcommand\bra[1]{\langle#1|}

\newcommand{\piJ}{\pi J}
\newcommand{\hf}{\tfrac{1}{2}}
\newcommand{\para}{\emph{para}\xspace}
\newcommand{\Para}{\emph{Para}\xspace}
\newcommand{\Htwo}{\ensuremath{\mathrm{H_2}}\xspace}
\newcommand{\TwoH}{\ensuremath{\mathrm{^2 H}}\xspace}
\newcommand{\Nfrtn}{\ensuremath{\mathrm{^{14}N}}\xspace}
\newcommand{\Bz}{B_z}

\newcommand{\Bbias}{B_\mathrm{bias}}
\newcommand{\BWOLF}{B_\mathrm{WOLF}}
\newcommand{\gI}{\gamma_I}
\newcommand{\gS}{\gamma_S}
\newcommand{\Ip}[1]{I^{+}_{#1}}
\renewcommand{\Im}[1]{I^{-}_{#1}}
\newcommand{\Iz}[1]{I_{{#1}z}}
\newcommand{\Sp}[1]{S^{+}_{#1}}
\newcommand{\Sm}[1]{S^{-}_{#1}}
\newcommand{\Sz}[1]{S_{{#1}z}}
\newcommand{\HA}{H_A}
\newcommand{\HB}{H_B}
\newcommand{\HC}{H_C}
\newcommand{\HD}{H_D}

\newcommand{\HWOLF}{H_\mathrm{WOLF}}
\newcommand{\boldI}{\boldsymbol{I}}
\newcommand{\wWOLF}{\omega_\mathrm{WOLF}}
\newcommand{\wST}{\omega_{ST}}

\newcommand{\wSTnut}{\omega^{ST}_\mathrm{nut}}

\begin{document}

\preprint{APS/123-QED}

\title{
Low-Frequency Excitation of Singlet-Triplet Transitions. \\Application to Nuclear Hyperpolarization
}

\author{Laurynas Dagys}
\author{Christian Bengs}
\author{Malcolm H. Levitt}

 \affiliation{Department of Chemistry, University of Southampton, SO17 1BJ, UK}%

\date{\today}

\begin{abstract}
Coupled pairs of nuclear spins-1/2 support one singlet state and three triplet states. Transitions between the singlet state and one of the triplet states may be driven by an oscillating low-frequency magnetic field, in the presence of couplings to a third nuclear spin, and a weak bias magnetic field. This phenomenon allows the generation of strong nuclear hyperpolarization of \Cth nuclei, starting from the nuclear singlet polarization of a \Hone spin pair,  associated with the enriched para spin isomer of hydrogen gas. Hyperpolarization is demonstrated for two molecular systems.

\end{abstract}

\maketitle


\section{\label{sec:intro}Introduction}
Magnetic resonance experiments usually involve the application of a strong magnetic field (typically, many Tesla) combined with radiofrequency pulses (for nuclear magnetic resonance, NMR) or microwave pulses (for electron spin resonance, ESR) that are resonant with the magnetic Zeeman transitions of the system. In these high-field conditions, the parts of the spin Hamiltonian that do not commute with the Zeeman Hamiltonian are usually removed. This \emph{secular approximation} leads to a major simplification of spin dynamical theory and is one of the cornerstones of modern NMR theory~\cite{ernst_principles_1987}. One consequence is that all transitions involving combinations or multiples of the Larmor frequencies are not observed in NMR or ESR spectra. Some exceptions to this paradigm exist, such as ``overtone" transitions in the NMR of nuclei with large quadrupolar moments~\cite{tycko_overtone_1987}, and the use of non-secular hyperfine couplings in solid-effect dynamic nuclear polarization~\cite{wenckebach_solid_2008}. 

In the case of NMR, the secular approximation breaks down in very small magnetic fields, such that the Larmor frequencies are comparable in magnitude to the spin-spin interactions~\cite{blanchard_measurement_2015}. Here we show that the mixing of states by non-secular spin-spin couplings in the low-field regime allows selected ``forbidden" transitions to be induced by oscillating magnetic fields. In suitable circumstances, this phenomenon allows the generation of strong hyperpolarization in the reaction products of hydrogen gas enriched in the \para spin isomer. We demonstrate the generation of two molecules in solution with high levels of \Cth polarization. One of these substances, fumarate, is a natural metabolite which has been used for the characterisation of cancer in magnetic resonance imaging (MRI)~\cite{hesketh_magnetic_2018}.

\begin{figure}[b]
\includegraphics[width=0.48\textwidth]{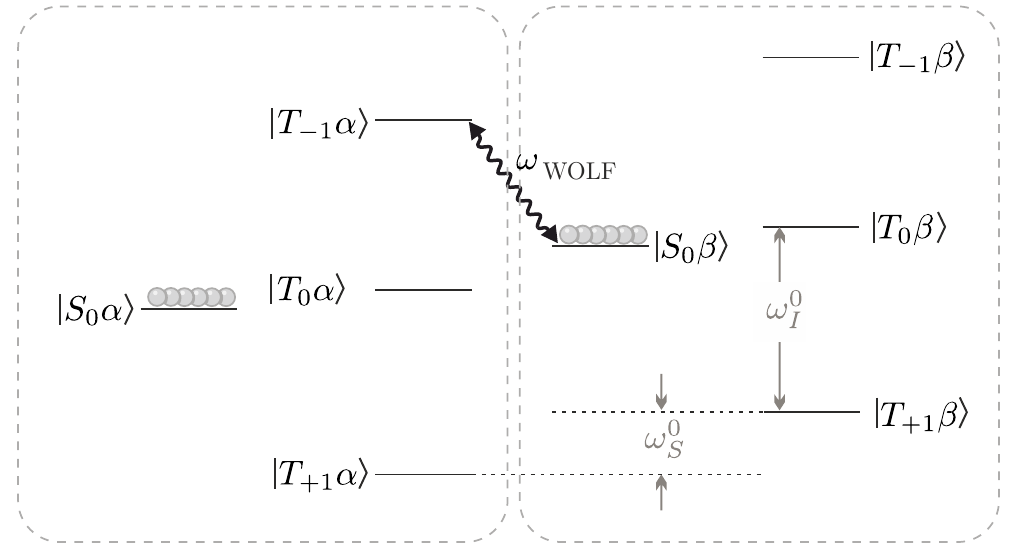}
\caption{\label{fig:energies} Approximate energy level diagram of a system of two $I$-spins and one $S$-spin, in the near-equivalence limit ($|J_{13}-J_{23}|\ll|J_{12}|$). The circles represent the population distribution for a fully populated singlet state between the two $I$-spins. The Larmor frequencies in the bias field $\Bbias$ are indicated: $\omega_I^0=-\gI\Bbias$ for the $I$-spins, and $\omega_S^0=-\gS\Bbias$ for the $S$-spins. The WOLF pulse is applied with a frequency $\wWOLF$ matching the transition frequency $\wST$ between the indicated pair of states (equation~\ref{eq:wST}). 
}
\end{figure}
\section{\label{sec:theory}Theory}
Consider an ensemble of nuclear three-spin-1/2 systems, each consisting of two nuclei $I_1$ and $I_2$ of one isotopic type with magnetogyric ratio $\gI$, and a third nucleus $S_3$ of a different isotopic type with magnetogyric ratio $\gS$. 
In the isotropic solution state, the three nuclei mutually interact by scalar spin-spin coupling terms mediated by the bonding electrons. These couplings consist of a homonuclear coupling $J_{12}$, and two heteronuclear couplings $J_{13}$ and $J_{23}$. We assume that the two heteronuclear couplings are unequal, $J_{13}\neq J_{23}$. 

\subsection{WOLF pulses}
A small magnetic field $\Bz$ is applied along the laboratory frame $z$-axis. This field is the sum of a time-independent ``bias" field $\Bbias$, and a time-dependent oscillating field denoted $\BWOLF$, such that
$\Bz(t)=\Bbias + \BWOLF(t)$. The oscillating field is given by
\begin{equation}
    \BWOLF(t)=\BWOLF^0 \cos(\wWOLF t),
\end{equation}
where $\wWOLF$ is the frequency of the oscillating field, and $\BWOLF^0$ is its peak amplitude. Since the applied oscillating field is small in magnitude and low in frequency, we refer to the oscillating magnetic field as a WOLF (\emph{Weak Oscillating Low Frequency}) pulse. 

Note that the $\Bbias$ and $\BWOLF$ fields are applied in the \emph{same direction}. This geometry differs from most NMR experiments, where oscillating fields are applied transverse to the main field, with few exceptions~\cite{tycko_overtone_1987,pileio_extremely_2009,sjolander_transition-selective_2016}.

If the magnetic fields are low enough that chemical shifts are negligible, the spin Hamiltonian may be written as a sum of five terms
\begin{equation}
\label{eq:Htot}
    H(t) = \HA + \HB + \HC + \HD+\HWOLF(t).
\end{equation}
These are given by
\begin{align}\label{eq:Hterms}
\HA &=
    - \Bbias
    \left(
    \gI (\Iz1+\Iz2)
    +\gS \Sz3
    \right)
\nonumber\\
 &+ 2\piJ_{12}\boldI_1\cdot\boldI_2
 +\pi(J_{13}+J_{23})(\Iz1+\Iz2)\Sz3,
\nonumber\\
\HB &=
    \pi(J_{13}-J_{23})(\Iz1-\Iz2)\Sz3,
\nonumber\\
\HC &=
    \hf\pi(J_{13}+J_{23})
\nonumber\\
   &\times (\Ip1\Sm3+\Ip2\Sm3
    +\Im1\Sp3+\Im2\Sp3
    ),
\nonumber\\
\HD &=
    \hf\pi(J_{13}-J_{23})
\nonumber\\
   &\times (\Ip1\Sm3-\Ip2\Sm3
    +\Im1\Sp3-\Im2\Sp3
    ),
\nonumber\\
    \HWOLF(t) & =
- \BWOLF(t) 
\left(
    \gI (\Iz1+\Iz2)
    +\gS \Sz3
    \right).
\end{align}
The terms $\HA$, $\HC$, and $\HWOLF$ are symmetric with respect to exchange of the two $I$-spins, while the terms $\HB$ and $\HD$ are antisymmetric. If the Larmor frequency of the nuclei in the bias field is larger than the spin-spin couplings, and the difference between the heteronuclear couplings is smaller than the homonuclear coupling ($|J_{13}-J_{23}|\ll |J_{12} |$, i.e. near-equivalence~\cite{tayler_singlet_2011}), then the $\HA$ term dominates. The terms $\HC$ and $\HWOLF$ commute with $\HA$ and may be regarded as secular, while the terms $\HB$ and $\HD$ do not commute with $\HA$ and are non-secular. The terms $\HB$ and $\HC$ both give rise to small eigenvalue shifts, which are neglected here, for the sake of brevity. The exchange-antisymmetric non-secular term $\HD$ is the most important one for the purposes of this paper. 

The eigenstates and eigenvalues of $\HA$ are sketched in figure~\ref{fig:energies}. The eigenstates are direct products of the singlet and triplet states of the $I$-spin pair with the Zeeman states of the $S$-spin, such that the triplet product states are symmetric and the singlet product states are anti-symmetric under permutation of spins 1 and 2:
\begin{equation}
\begin{aligned}
    (12)\ket{T_{M}m_S}&=(+1)\ket{T_{0}m_S},
    \\
    (12)\ket{S_{0}m_S}&=(-1)\ket{S_{0}m_S},
\end{aligned}    
\end{equation}
where $M\in\{-1,0,1\}$ and the Zeeman eigenequations are
\begin{equation}
\begin{aligned}
(I_{1z}+I_{2z})\ket{T_{M}m_S}
     &= M\ket{T_{M}m_S}, \\
(I_{1z}+I_{2z})\ket{S_{0}m_S}
     &= 0, \\
S_{3z}\ket{T_{M}m_S} 
    &=m_S\ket{T_{M}m_S}, \\
S_{3z}\ket{S_{0}m_S} 
    &=m_S\ket{S_{0}m_S}.
\end{aligned}
\end{equation}
The symbols $\alpha$ and $\beta$ are used to denoted $m_S=\pm\hf$ respectively. 

Consider the ``forbidden" transition between the $\ket{S_0\beta}$ and $\ket{T_{-1}\alpha}$ state, indicated in figure~\ref{fig:energies}. All Hamiltonian terms in equation~\ref{eq:Hterms} have a zero matrix element connecting these two states, except for the non-secular antisymmetric term $\HD$, with matrix elements given by
\begin{equation}
    \bra{S_0\beta}\HD\ket{T_{-1}\alpha} 
     = 
    \bra{T_{-1}\alpha}\HD\ket{S_0\beta} 
     = 2^{-1/2} \pi (J_{13}-J_{23}).
\end{equation}
The difference between the corresponding diagonal elements of $\HA$ is given by
%
\begin{equation}\label{eq:wST}
    \begin{aligned}
\wST&=\bra{T_{-1}\alpha}H_{0}\ket{T_{-1}\alpha}-\bra{S_{0}\beta}H_{0}\ket{S_{0}\beta}
\\
&=\Bbias(\gI-\gS)+\frac{\pi}{2}(4J_{12}-J_{13}-J_{23}),
    \end{aligned}
\end{equation}
where the notation $\wST$ indicates ``singlet-triplet transition". Note that this transition frequency includes a combination of $I$-spin and $S$-spin Larmor frequencies. 

The term $\HD$ induces a slight mixing of these two $\HA$ eigenstates.
The small degree of state mixing would have very little effect, if it were not for the time-dependence introduced by the coupling to the oscillating magnetic field, $\HWOLF(t)$. If the oscillation frequency matches the singlet-triplet transition frequency ($|\wWOLF|\simeq|\wST|$), the  periodic time-dependence drives coherent transitions between these two eigenstates. 
For example, suppose the initial state of the system consists of a strongly populated state $\ket{S_0\beta}$, and a completely depleted state $\ket{T_{-1}\alpha}$. The density operator is described by
\begin{align}\label{eq:popsbeforeWOLF}
    \bra{S_0\beta}\,\rho(0)\,\ket{S_0\beta} &=1,
\nonumber\\
    \bra{T_{-1}\alpha}\,\rho(0)\,\ket{T_{-1}\alpha} &=0.
\end{align}
Consider a WOLF pulse applied for a duration $\tau$, on the resonance condition $\wWOLF=\wST$. As shown in the Supplemental Material~\cite{noauthor_see_nodate}, the spin-state populations after the pulse are approximately given by
\begin{align}\label{eq:popsafterWOLF}
    \bra{S_0\beta}\,\rho(\tau)\,\ket{S_0\beta} &\simeq
        \hf\left(1 + \cos(\wSTnut \tau)
        \right),
\nonumber\\
    \bra{T_{-1}\alpha}\,\rho(\tau)\,\ket{T_{-1}\alpha} &\simeq
        \hf\left(1 - \cos(\wSTnut \tau)
        \right),
\end{align}
where the singlet-triplet nutation frequency under the WOLF pulse is given by
\begin{equation}\label{eq:wSTnut}
    \begin{aligned}
\wSTnut=2\pi\sqrt{J^{2}_{13}+J^2_{23}}\sin(\phi+\theta/2)J_{1}(A).
    \end{aligned}
\end{equation}

The angles $\theta$ and $\phi$ are given by
\begin{equation}
\begin{aligned}
\theta &=\arctantwo(2 J_{12},J_{13}-J_{23}),
\\
\phi&=\arctantwo(J_{13}+J_{23},J_{13}-J_{23}),
\end{aligned}
\end{equation}
such that $\theta,\phi\ll1$ in the near-equivalence regime. 

The symbol $J_{1}$ denotes a Bessel function of the first kind with its argument given by 
\begin{equation}
    A=(\gI-\gS)\BWOLF^0/\wST.
\end{equation}
These equations apply approximately in the limits \mbox{$|J_{13}-J_{23}|\ll|J_{12}|$} and $J^{2}_{13}+J^{2}_{23} \ll \wST^2 $.

It is therefore possible to transport population completely from one state to the other by applying a WOLF pulse of duration 
$\pi/\wSTnut$. The inversion speed is maximised by choosing the peak amplitude of the WOLF pulse to equal $\BWOLF^{0}\simeq 2 \Bbias$, at which point the Bessel function $J_{1}(A)$ reaches its approximate maximum.
Since there are practical limits on the generation of large oscillating magnetic fields, this technique is most appropriate for low-field magnetic resonance. 

The use of oscillating magnetic fields for the selective excitation of transitions has previously been explored in the context of ultralow-field NMR~\cite{sjolander_transition-selective_2016}. However, those experiments follow the familiar paradigm of resonant excitation, in which transitions are induced by modulating off-diagonal Hamiltonian terms at a frequency that matches the transition energy. In the current case, on the other hand, the off-diagonal non-secular terms are time-independent, and it is the diagonal terms that are given a periodic time-dependence by the oscillating applied field. There is a distant relationship with selective excitation in magic-angle-spinning solid-state NMR~\cite{caravatti_selective_1983}.

\subsection{Parahydrogen-induced polarization}
The resonant driving of singlet-triplet transitions is particularly useful in the context of parahydrogen-induced polarization (PHIP), a technique which is widely used to enhance NMR signals~\cite{bowers_transformation_1986,bowers_parahydrogen_1987,goldman_conversion_2005,adams_reversible_2009,kadlecek_optimal_2010,pravdivtsev_highly_2014,theis_light-sabre_2014,theis_microtesla_2015,ripka_hyperpolarized_2018,eills_singlet_2017,eills_polarization_2019,barskiy_sabre_2019,birchall_quantifying_2020,knecht_rapid_2021,eriksson_improving_2021}. In this method, hydrogen gas is enriched in the \para spin isomer and reacted with a substrate in the presence of a catalyst. The proton pair of the product molecule exhibits excess population in the singlet state. The strongly enhanced singlet spin order is converted into hyperpolarized magnetization of heteronuclei such as \Cth by applying a sequence of magnetic fields. A range of suitable techniques has been developed~\cite{johannesson_transfer_2004,goldman_conversion_2005,kadlecek_optimal_2010,pravdivtsev_highly_2014,theis_light-sabre_2014,theis_microtesla_2015,eills_singlet_2017,barskiy_sabre_2019,eills_polarization_2019,bengs_robust_2020}.

We now show that the application of a WOLF pulse in a small bias magnetic field leads to strong hyperpolarization of \Cth nuclei in the reaction products of \para-enriched \Htwo gas, with potential advantages over other methods, as discussed below. 

The principle of the experiment is shown in figure~\ref{fig:energies}. The spin state populations (indicated by balls) are given for the case that the spins $I_1$ and $I_2$ are protons originating from the \para-enriched hydrogen and the spin $S_3$ is a \Cth nucleus in the product molecule. The populations of the $\ket{S_0\alpha}$ and $\ket{S_0\beta}$ states are strongly enhanced due to their provenance as the nuclei of the \para-enriched \Htwo spin isomer. Since these two populations are equal there is no polarization of the $S$-spin at this stage. However, if the WOLF pulse transports the population from the $\ket{S_0\beta}$ state to the $\ket{T_{-}\alpha}$ state, as shown in equation~\ref{eq:popsafterWOLF}, the resulting population distribution has excess population in the $\ket{S_0\alpha}$ and $\ket{T_{-}\alpha}$ states. Since the $S$-spin is in the state $\ket{\alpha}$ in both cases, this corresponds to a high degree of \Cth polarization. A \Cth polarization of the order of unity represents an enhancement of the \Cth NMR signals by around 5 orders of magnitude, relative to ordinary NMR based on thermal equilibrium polarization in a strong magnetic field.

We propose the acronym WEREWOLF (Whopping Enhancement Requires Excitation by Weak Oscillating Low Fields) for \para-hydrogen-induced polarization of heteronuclei using WOLF pulses for the singlet-triplet population transfer.

\begin{figure}[h]
\includegraphics[width=0.42\textwidth]{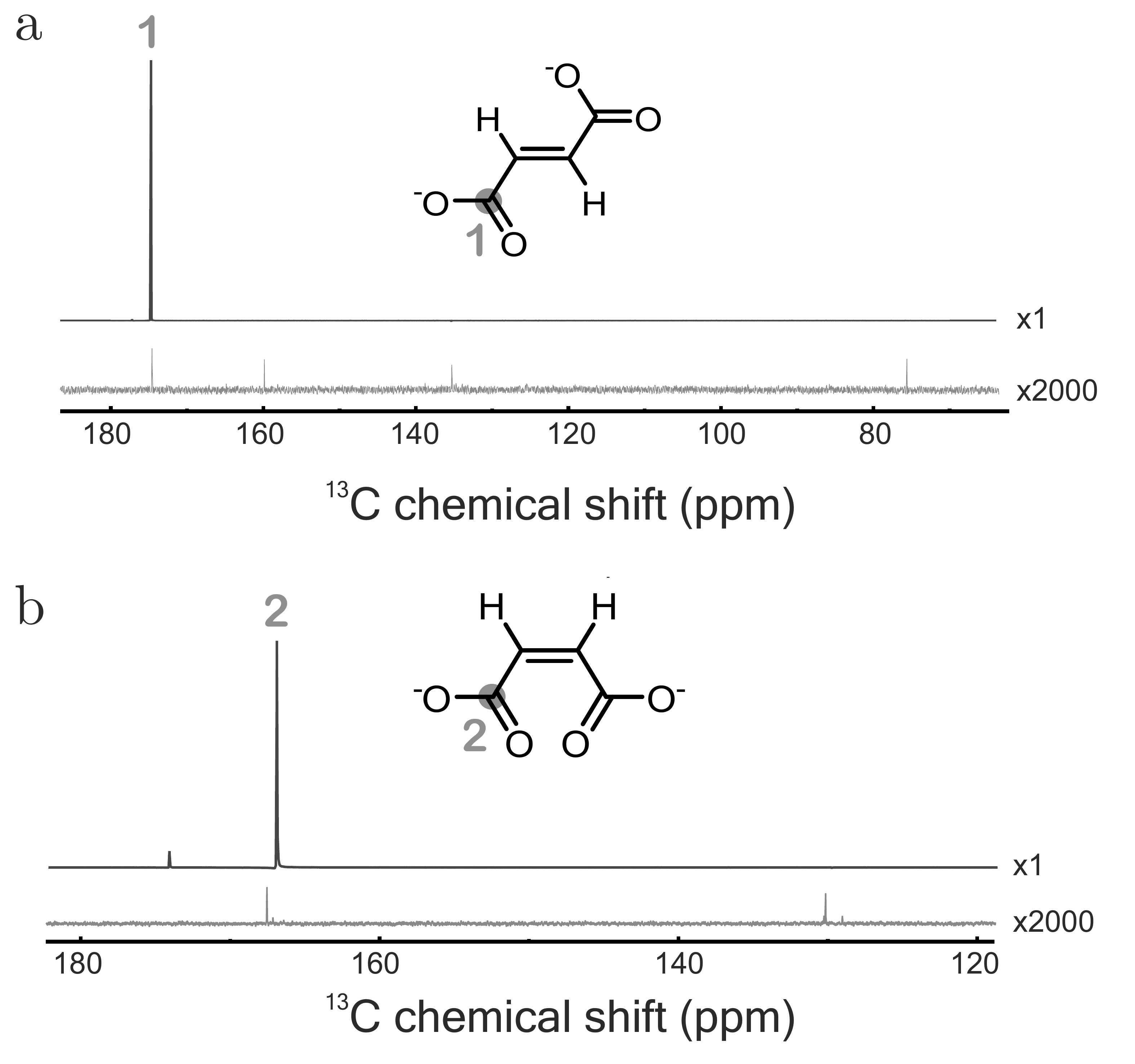}
\caption{\label{fig:spectra} \Hone-decoupled \Cth spectra of (a) fumarate and (b) maleate at a field of 9.41~T. Single-transient WEREWOLF-hyperpolarized \Cth spectra are compared with conventional \Cth NMR spectra acquired at thermal equilibrium, averaged over 360 transients (for fumarate) or 512 transients (for maleate). The strong \Cth peaks are from hyperpolarized naturally occurring \Cth nuclei at the molecular sites indicated by filled circles. The \Cth peaks at 160~ppm and 76~ppm in (a) correspond to unreacted disodium acetylene dicarboxylate. The small signal at 172~ppm in (b) is  attributed to succinate generated by secondary hydrogenation.
}
\end{figure}
\begin{figure*}[htb]
\includegraphics[width=0.8\textwidth]{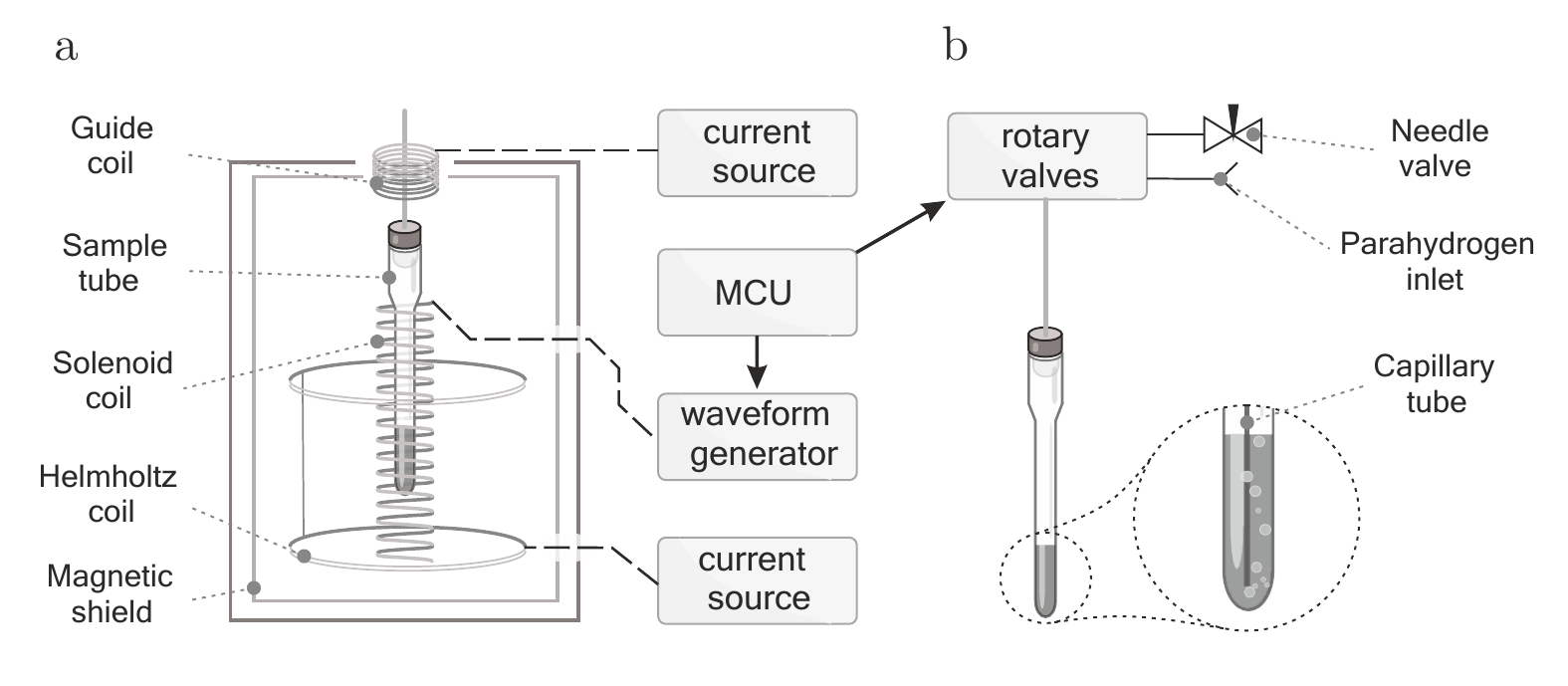}
\caption{\label{fig:setup} Schematic diagram of the experimental setup. (a) Mu-metal shield and associated components. During the WOLF pulse, the Helmholtz coil generates the bias field $\Bbias$ whereas the solenoid coil produces the oscillating field $\BWOLF(t)$ (b) Gas-handling apparatus including a thick-wall NMR tube equipped with a capillary for the \para-enriched \Htwo gas. MCU - micro-controller unit. More details are given in the Supporting Information~\cite{noauthor_see_nodate}.
}
\end{figure*}
\begin{table}[bth]
\caption{\label{tab:couplings}%
Spin-spin coupling parameters for (1-\Cth)fumarate and (1-\Cth)maleate, adapted from reference~\cite{bengs_robust_2020}. 
}
\begin{ruledtabular}
\begin{tabular}{lccc}
Compound        &$J_{12}$(Hz)   &$J_{23}$(Hz)   &$J_{13}$(Hz)\\
\hline
Fumarate    &15.9           &5.8            &3.3        \\
Maleate         &12.3           &12.9           &2.5        \\
\end{tabular}
\end{ruledtabular}
\end{table}

\section{\label{sec:Methods}Methods}
\subsection{\label{sec:Materials}Materials}

The substances fumarate (\emph{E}-butenedioate) and  maleate (\emph{Z}-butenedioate) were used for demonstrations of WEREWOLF. The chemical structures are shown in figure~\ref{fig:spectra}.
Both compounds are formed by the catalytic hydrogenation of acetylene dicarboxylate, using a ruthenium-based catalyst for fumarate~\cite{ripka_hyperpolarized_2018}, and a rhodium-based catalyst for maleate~\cite{eills_polarization_2019}. Chemical details are given in the Supplemental Material~\cite{noauthor_see_nodate}. 

About 2\% of fumarate and maleate molecules contain a naturally-occurring \Cth nucleus at the sites shown in figure~\ref{fig:spectra}. The two \Hone nuclei and the \Cth nucleus form a three-spin-1/2 system of the type discussed above. The $J$-coupling parameters for the two molecular systems are given in table~\ref{tab:couplings}.

\Para-enriched hydrogen was  produced by passing hydrogen gas over an iron oxide catalyst cooled by liquid nitrogen. 
\subsection{\label{sec:Equipment}Equipment}
A sketch of the equipment is shown in figure~\ref{fig:setup}. This consists of a mu-metal chamber for magnetic shielding, equipped with a solenoid coil (for the WOLF pulses) and a Helmholtz coil (for the bias field). A guide coil allows the sample to traverse the chamber walls without passing through zero magnetic field. 
The coils are driven by waveform generator and stable current sources. The equipment set also includes computer-actuated valves and an NMR tube equipped with a capillary, allowing bubbling of 
\para-enriched \Htwo under microprocessor control. 
Full details are in the Supplemental Material~\cite{noauthor_see_nodate}. 
\begin{figure}[h]
\includegraphics[width=0.48\textwidth]{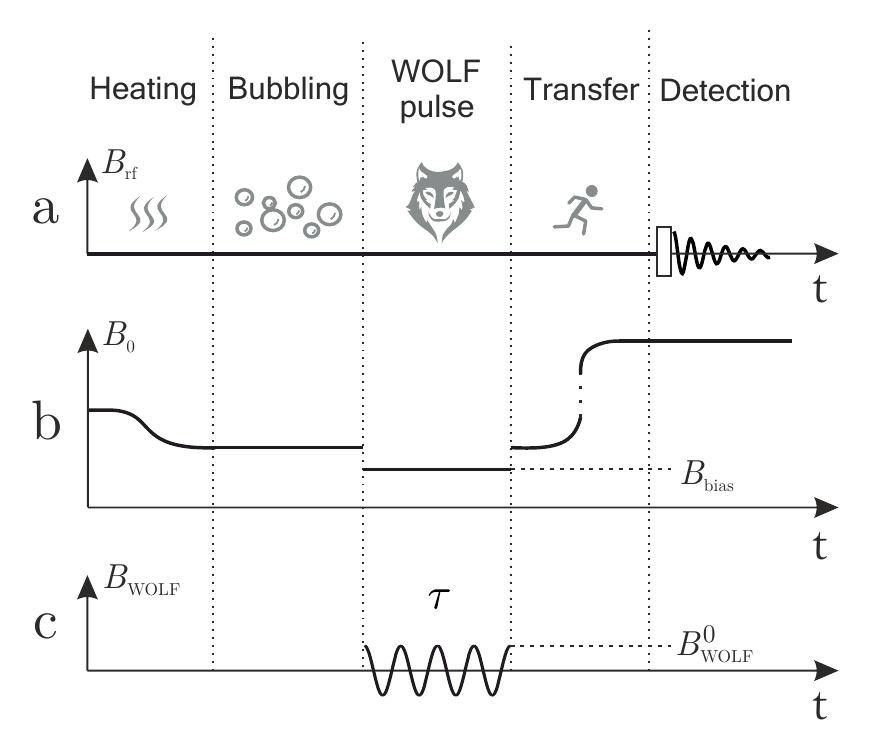}
\caption{\label{fig:sequence} 
Timing diagram for the WEREWOLF procedure, involving bubbling of the sample with \para-enriched hydrogen gas and a sequence of magnetic fields. 
(a) \Cth radio-frequency pulse applied in high magnetic field at the end of the procedure and NMR signal acquisition. 
(b) Magnetic field along the $z$-axis, showing the ambient laboratory field during sample heating, the reduction in field as the sample is placed in the shield, the bias field $\Bbias$ during the WOLF pulse, the removal of the sample from the shield and insertion into the high-field NMR magnet. 
(c) Oscillating WOLF pulse field, applied for a duration $\tau$, with a frequency $\wWOLF$ and peak amplitude $\BWOLF^0$. The total field experienced by the sample is the sum of (b) and (c). 
}
\end{figure}
\subsection{\label{sec:Procedure}Experimental Procedure}
Figure~\ref{fig:sequence} shows a timing diagram of the WEREWOLF experiment, showing the magnetic fields experienced by the sample as a function of time. Each experiment starts by warming 
the sample mixture to $\sim90 \degC$ in the ambient magnetic field of the laboratory ($\sim$ 110~$\mu$T), followed by insertion into the magnetic shield. \Para-enriched hydrogen gas is bubbled through the solution at 6 bar pressure. The bias field is reduced to $\Bbias=2\uT$ while the oscillating WOLF pulse is applied with amplitude $\BWOLF^0=2\uT$ for an interval $\tau$. The sample is removed from the shield and inserted by hand into the 9.41~T NMR magnet where a 90\deg \Cth radio-frequency is applied. The \Cth free-induction decay is acquired in the presence of \Hone decoupling. Fourier transformation of the NMR signal gives the \Cth NMR spectrum. Full details are in the Supplemental Material~\cite{noauthor_see_nodate}. 

\begin{figure}[h]
\includegraphics[width=0.4\textwidth]{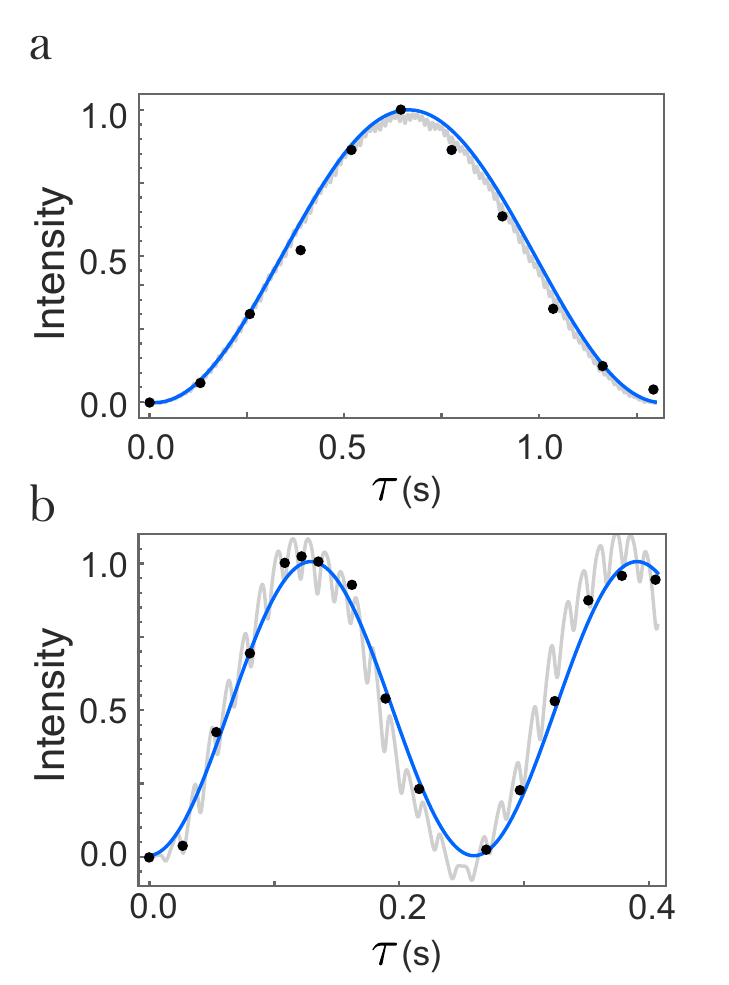}
\caption{\label{fig:cycles} Hyperpolarized fumarate (a) and maleate (b) \Cth signal intensities as a function of the WOLF pulse duration $\tau$. 
The WOLF pulse frequencies are 77.3~Hz for fumarate and 74.0~Hz for maleate. The black dots depict integrated experimental signal amplitudes, normalized to $1$ for the maximum value. The grey lines represent numerical \emph{SpinDynamica}~\cite{bengs_spindynamica_2018} simulations for the coupling parameters in table~\ref{tab:couplings}. The blue lines show analytical solutions given by equation~\ref{eq:popsafterWOLF}. All values of $\tau$ used for the experiments are integer numbers of the WOLF period $2\pi/\wWOLF$, which is 12.94~ms for fumarate and 13.51~ms for maleate.
}
\end{figure}
\section{\label{sec:results}Results}
Figure~\ref{fig:spectra} shows single-transient hyperpolarized \Cth NMR spectra for (a) fumarate and (b) maleate, obtained using the WEREWOLF procedure. The WOLF pulse parameters were $\wWOLF/2\pi=77.3\Hertz$ and $\tau=0.65\sec$ for fumarate, and $\wWOLF/2\pi=74.0\Hertz$ and $\tau=0.12\sec$ for maleate. 

Conventional \Cth NMR spectra obtained by multiple signal acquisitions on the hydrogenated samples after thermal equilibration are also shown in figure~\ref{fig:spectra}. Comparison of these spectra allows the estimation of the \Cth polarization levels achieved by WEREWOLF, which are $p\simeq8\%$ for fumarate and $p\simeq19\%$ for maleate. These results are highly competitive with previous work~\cite{rodin_constant-adiabaticity_2021,rodin_constant-adiabaticity_2021-1,knecht_rapid_2021}, especially when the relatively low \para-hydrogen enrichment levels, crude apparatus, manual sample transport, and sub-optimal reaction conditions are taken into account.

Integrated \Cth signal amplitudes as a function of the WOLF pulse duration $\tau$ are shown in figure~\ref{fig:cycles}. The coherent oscillations of the hyperpolarized magnetization are striking. Each experimental point was obtained from a separate experiment on a fresh sample. The normalized experimental data is compared with numerical simulations and analytical curves derived from equations~\ref{eq:popsafterWOLF} and \ref{eq:wSTnut}. The agreement between both curves and the experimental data is gratifying.

\section{\label{sec:Conclusions}Conclusions}
Magnetic resonance phenomena involving non-secular spin-spin couplings are normally encountered in systems with large nuclear quadrupolar couplings~\cite{tycko_overtone_1987} or hyperfine couplings to unpaired electrons, as in solid-effect dynamic nuclear polarization (DNP)~\cite{wenckebach_solid_2008}. The work described here shows that non-secular couplings may also be exploited in purely nuclear spin systems, albeit at an energy scale which is $\sim9$ orders of magnitude lower than in solid-effect DNP. Here too, non-secular effects allow ``forbidden" transitions to couple to the electromagnetic field, allowing the generation of strong nuclear hyperpolarization.

There are several other techniques for performing the transformation of nuclear singlet order to heteronuclear magnetization, including pulse techniques in high field~\cite{goldman_conversion_2005,kadlecek_optimal_2010,theis_light-sabre_2014,eills_singlet_2017,bengs_robust_2020}, field-cycling and level-anticrossing phenomena ~\cite{johannesson_transfer_2004,pravdivtsev_highly_2014,theis_microtesla_2015,eills_polarization_2019},
and low-field methods~\cite{theis_microtesla_2015,eills_polarization_2019,eriksson_improving_2021}.
The WOLF method described here is highly intuitive, provides a very fast exchange of singlet and triplet populations, and has further potential advantages over current methods. Since the method works in the presence of a bias magnetic field, problems associated with ultralow-field NMR are avoided, such as short coherence lifetimes in the presence of rapidly relaxing nuclear species such as \TwoH and \Nfrtn~\cite{birchall_quantifying_2020}. 
The bias field resolves the Larmor frequencies of the different nuclear isotopes, making it possible to implement heteronuclear spin decoupling by resonant transverse oscillating fields. Furthermore, the coherent oscillatory spin evolution under the WOLF pulses suggests that the full palette of ``pulse tricks" developed for conventional high-field NMR may be deployed in this regime, including error-compensating composite pulses~\cite{levitt_composite_1986} and adiabatic frequency sweeps~\cite{baum_broadband_1985}. Applications are envisaged to other hyperpolarization techniques, such as the SABRE (Signal Amplification by Reversible Exchange) method~\cite{adams_reversible_2009,barskiy_sabre_2019}. The method described in the current paper might be related to a recently described variation of SABRE~\cite{eriksson_improving_2021}.

\begin{acknowledgments}
We acknowledge funding received by the Marie Skłodowska-Curie program of the European Union (grant number 766402), the European Research Council (grant 786707-FunMagResBeacons), and EPSRC-UK (grants EP/P009980/1, EP/P030491/1, EP/V055593/1). We thank Soumya Singha Roy for discussions.
\end{acknowledgments}

\bibliography{References/LD&CB-WEREWOLF}%

\end{document}


\title{Supplemental Material for Low-Frequency Excitation of Singlet-Triplet Transitions. \\Application to Nuclear Hyperpolarization
}

\author{Laurynas Dagys}
\author{Christian Bengs}
\author{Malcolm H. Levitt}

\affiliation{Department of Chemistry, University of Southampton, SO17 1BJ, UK}

\date{\today}

\maketitle

\section{WOLF pulse dynamics}\label{seq:dynamics}

\begin{figure}[H]
\centering
\includegraphics[width=0.42\textwidth]{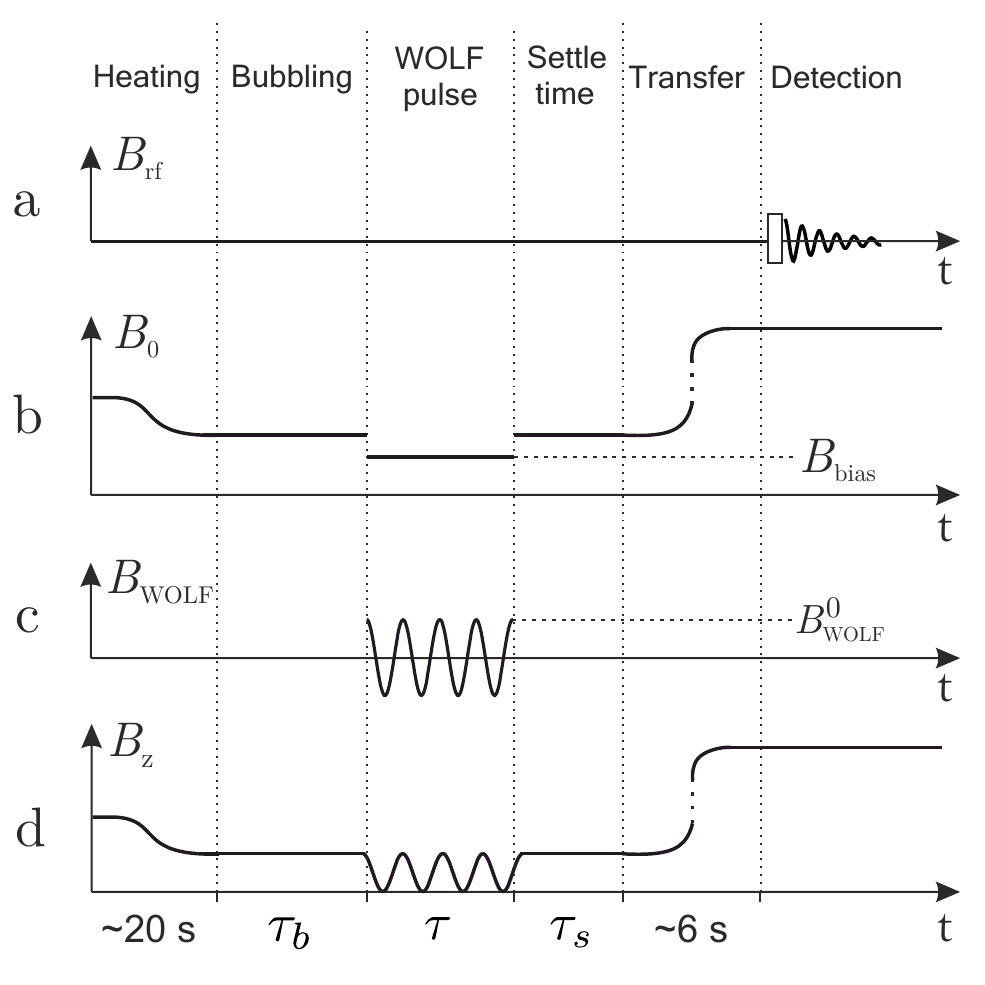}
\caption{\label{fig:sequence_SI} 
Detailed timing diagram for the WEREWOLF procedure. 
(a) Radiofrequency \Cth pulse applied in high magnetic field at the end of the procedure and NMR signal acquisition. 
(b) Magnetic field along the z-axis, showing the ambient laboratory field, the reduction in field as the sample is placed in the shield, the bias field $\Bbias$ during the WOLF pulse, the removal of the sample from the shield and insertion into the high-field NMR magnet.
(c) Oscillating WOLF pulse field, applied for a duration $\tau$, with a frequency $\wWOLF$ and peak amplitude $\BWOLF^0$. (d) The total field experienced by the sample is the sum of (b) and (c). 
Timings for separate phases of the sequence are given below, where bubbling time $\tau_b$ was set to 30~s (for fumarate) and 10~s (for maleate). The time period $(\tau+\tau_{s})$ was set to 5~s for all experiments.
}
\end{figure}

The complete Hamiltonian given by equation 2 in the letter preserves the total $z$-angular momentum of the system
\begin{equation}
    \begin{aligned}
    {[}H(t),I_{1z}+I_{2z}+S_{3z}{]}=0.
    \end{aligned}
\end{equation}
This suggests that the basis states highlighted in figure 1 separate the matrix representation of $H(t)$ into two $3\times3$ and two $1\times1$ blocks. The $1\times1$ blocks may be disregarded, the $3\times3$ blocks are generated by the sets
\begin{equation}
    \begin{aligned}
    V&=\{\ket{S_{0}\alpha},\ket{T_{0}\alpha},\ket{T_{+1}\beta}\},
    \\
    W&=\{\ket{S_{0}\beta},\ket{T_{0}\beta},\ket{T_{-1}\alpha}\}.
    \end{aligned}
\end{equation}
The standard WEREWOLF experiment (Fig.~\ref{fig:sequence_SI}) aims to create positive heteronuclear magnetisation which may be achieved through a population swap of states $\ket{S_{0}\beta}$ and $\ket{T_{-1}\alpha}$. We thus consider the restriction of $H(t)$ to $W$, which may be explicitly given as follows
\begin{equation}
    \begin{aligned}
&{[}H(t){]}_{W}=
\\
&\left[\begin{array}{ccc}
\frac{1}{2} (B(t) \gS-3\pi J_{12})
& 
\frac{\pi}{2}(J_{23}-J_{13}) 
&
\frac{\pi}{\sqrt{2}}(J_{13}-J_{23})
\\
\frac{\pi}{2}(J_{23}-J_{13}) & \frac{1}{2} (B(t) \gS+\pi J_{12})  & \frac{\pi}{\sqrt{2}}(J_{13}+J_{23})
\\
\frac{\pi}{\sqrt{2}}(J_{13}-J_{23}) 
& 
\frac{\pi}{\sqrt{2}}(J_{13}+J_{23})
& 
(\gI-\frac{1}{2}\gS)B(t)+\frac{\pi}{2}(J_{12}-J_{13}-J_{23})
\end{array}\right],
    \end{aligned}
\end{equation}
where
\begin{equation}
    \begin{aligned}
B(t)=\Bbias+\BWOLF^{0}\cos(\wWOLF t).
    \end{aligned}
\end{equation}

Although small, the mixing between the states $\ket{S_{0}\alpha}$ and $\ket{T_{0}\alpha}$ is not negligible during the WOLF pulse dynamics. To account for the mixing we define a set of rotated basis states 
\begin{equation}
    \begin{aligned}
    W_{\theta}
    =&
    \{
    \ket{S^{'}_{0}\beta},
    \ket{T^{'}_{0}\beta},
    \ket{T^{'}_{-1}\alpha}
    \}
    \\
    =&
    \{\cos(\theta/2)\ket{S_{0}\beta}+\sin(\theta/2)\ket{T_{0}\beta},\cos(\theta/2)\ket{T_{0}\beta}-\sin(\theta/2)\ket{S_{0}\beta},
    \ket{T_{-1}\alpha}
    \},
    \end{aligned}
\end{equation}
parametrised by the angle $\theta$. The angle $\theta$ is chosen to satisfy 
\begin{equation}
    \begin{aligned}
    \theta=\arctantwo(2 J_{12},J_{13}-J_{23}).
    \end{aligned}
\end{equation}
In the near-equivalence regime the matrix representation of $H(t)$ restricted to $W_{\theta}$ is given by
\begin{equation}
\label{eq:tilted_mat}
    \begin{aligned}
&{[}H(t){]}_{W_{\theta}}\simeq
\\
&\left[\begin{array}{ccc}
\frac{1}{2} (B(t) \gS-3\pi J_{12})
& 
0
&
\frac{\pi}{\sqrt{2}}(\cos(\theta/2)(J_{13}-J_{23})+\sin(\theta/2)(J_{13}+J_{23}))
\\
0
& 
\frac{1}{2} (B(t) \gS+\pi J_{12})  
& 
\frac{\pi}{\sqrt{2}}(\cos(\theta/2)(J_{13}+J_{23})-\sin(\theta/2)(J_{13}-J_{23}))
\\
\blacksquare
& 
\blacksquare
& 
(\gI-\frac{1}{2}\gS)B(t)+\frac{\pi}{2}(J_{12}-J_{13}-J_{23})
\end{array}\right],
    \end{aligned}
\end{equation}
where the black squares indicate the fact that ${[}H(t){]}_{W_{\theta}}$ is equal to its transpose.

According to equation \ref{eq:tilted_mat} the transition frequency between states $\ket{S^{'}_{0}\beta}$ and $\ket{T^{'}_{-1}\alpha}$ in the near-equivalence regime is given by
\begin{equation}\label{eq:omega_st}
    \begin{aligned}
\wST&=\bra{T^{'}_{-1}\alpha}H_{0}\ket{T^{'}_{-1}\alpha}-\bra{S^{'}_{0}\beta}H_{0}\ket{S^{'}_{0}\beta}
\\
&=\Bbias(\gI-\gS)+\frac{\pi}{2}(4J_{12}-J_{13}-J_{23}),
    \end{aligned}
\end{equation}
whereas the transition frequency between $\ket{T^{'}_{0}\beta}$ and $\ket{T^{'}_{-1}\alpha}$ is given by
\begin{equation}
    \begin{aligned}
\wTT&=\bra{T^{'}_{-1}\alpha}H_{0}\ket{T^{'}_{-1}\alpha}-\bra{S^{'}_{0}\beta}H_{0}\ket{S^{'}_{0}\beta}
\\
&=\Bbias(\gI-\gS)-\frac{\pi}{2}(J_{13}+J_{23}).
    \end{aligned}
\end{equation}
Application of a WOLF pulse with $\wWOLF=\wST$ causes a resonant modulation of the energy difference between states $\ket{S^{'}_{0}\beta}$ and $\ket{T^{'}_{-1}\alpha}$, but causes an off-resonant modulation of the energy difference between states $\ket{T^{'}_{0}\beta}$ and $\ket{T^{'}_{-1}\alpha}$ on the order of $\sim J_{12}$. To a first approximation we may neglect the state $\ket{T^{'}_{0}\beta}$ altogether, and consider a fictitious two-level system evolving under the Hamiltonian $h(t)$
\begin{equation}
    \begin{aligned}
h(t)=
&\left[\begin{array}{cc}
     \bra{S^{'}_{0}\beta}H(t)\ket{S^{'}_{0}\beta}
     &  
     \bra{S^{'}_{0}\beta}H(t)\ket{T^{'}_{-1}\alpha}
     \\
     \bra{T^{'}_{-1}\alpha}H(t)\ket{S^{'}_{0}\beta}
     &
     \bra{T^{'}_{-1}}H(t)\ket{T^{'}_{-1}\alpha}
\end{array}\right]
\\
=&\left[\begin{array}{cc}
     \frac{1}{2} (B(t) \gS-3\pi J_{12})
     &  
     \frac{\pi}{\sqrt{2}}(\cos(\theta/2)(J_{13}-J_{23})+\sin(\theta/2)(J_{13}+J_{23}))
     \\
     \frac{\pi}{\sqrt{2}}(\cos(\theta/2)(J_{13}-J_{23})+\sin(\theta/2)(J_{13}+J_{23}))
     &
     (\gI-\frac{1}{2}\gS)B(t)+\frac{\pi}{2}(J_{12}-J_{13}-J_{23})
\end{array}\right].
    \end{aligned}
\end{equation}
In terms of {\em normalised} Pauli matrices $(\sigma_{j}/2)$
the Hamiltonian $h(t)$ may be expressed as shown below
\begin{equation}
    \begin{aligned}
h(t)=\omega_{0}\sigma_{0}+\omega_{x}\sigma_{x}/2+\omega_{z}(t)\sigma_{z}/2,
    \end{aligned}
\end{equation}
with 
\begin{equation}
    \begin{aligned}
&\omega_{0}=
\frac{1}{2}\gI(\Bbias+\BWOLF^0\cos(\wST t))-\frac{\pi}{4}(2J_{12}+J_{13}+J_{23}),
\\
&\omega_{x}=\pi\sqrt{2}(\cos(\theta/2)(J_{13}-J_{23})+\sin(\theta/2)(J_{13}+J_{23})),
\\
&\omega_{z}(t)=-\BWOLF^0(\gI-\gS)\cos(\wST t)+\Bbias(\gS-\gI)+\frac{\pi}{2}(J_{13}+J_{23}-4J_{12})
\\
&\phantom{\omega_{z}(t)}=-\BWOLF^0(\gI-\gS)\cos(\wST t)-\wST.
    \end{aligned}
\end{equation}
The Pauli coefficient $\omega_{x}$ may alternatively be expressed as follows
\begin{equation}
    \begin{aligned}
&\omega_{x}=
2\pi\sqrt{J^{2}_{13}+J^2_{23}}\sin(\phi+\theta/2),
\\
&\phi=\arctantwo(J_{13}+J_{23},J_{13}-J_{23}),
    \end{aligned}
\end{equation}
where we made use of the following relations
\begin{equation}
    \begin{aligned}
\sqrt{J_{13}+J_{23}}\sin(\phi)=(J_{13}-J_{23})/\sqrt{2},
\\
\sqrt{J_{13}+J_{23}}\cos(\phi)=(J_{13}+J_{23})/\sqrt{2}.
    \end{aligned}
\end{equation}

Within our approximations the $\omega_{0}$-term introduces an overall phase shift and may be discarded. The WOLF pulse dynamics may now be approximately described within a ``jolting'' interaction frame. The jolting frame represents a rotating frame with a time-dependent rotation frequency. The time-dependent rotation frequency and the rotation angle $\psi(t)$ are related as follows
\begin{equation}
    \begin{aligned}
\psi(t)=\int_{0}^{t}\omega_{z}(s)ds.
    \end{aligned}
\end{equation}
The jolting frame Hamiltonian $\tilde{h}(t)$ is then given by
\begin{equation}
    \begin{aligned}
\tilde{h}(t)
=&\frac{\omega_{x}}{2}\exp(+\iu \psi(t)\sigma_{z}/2)\sigma_{x}\exp(-\iu \psi(t)\sigma_{z}/2)
\\
=&\frac{\omega_{x}}{4}(\exp(+\iu \psi(t))\sigma_{+}+\exp(-\iu \psi(t))\sigma_{-}).
    \end{aligned}
\end{equation}
We may expand the jolting frame Hamiltonian as a Fourier series
\begin{equation}
    \begin{aligned}
\tilde{h}(t)=&\frac{\omega_{x}}{4}
\bigg(
\sum_{n}J_{n}(A)\exp(-\iu (n-1) \wST t)\sigma_{+}
+\sum_{n}J_{-n}(A)\exp(-\iu (n+1) \wST t)\sigma_{-}\bigg),
    \end{aligned}
\end{equation}
where $J_{n}(x)$ is the $n$'th Bessel function of the first kind and the argument $A$ is defined as follows
\begin{equation}
    \begin{aligned}
A=(\gI-\gS)\BWOLF^0/\wST.
    \end{aligned}
\end{equation}
For $\vert \omega_{x}/\wST\vert \ll 1$, which is bounded from above by
\begin{equation}
    \begin{aligned}
\vert \omega_{x}/\wST\vert\leq2\pi\sqrt{J^{2}_{13}+J^{2}_{23}}/\vert\wST\vert,
    \end{aligned}
\end{equation}
we may neglect the time-dependent terms of $\tilde{h}(t)$. This approach is equivalent to truncating the average Hamiltonian after first order
\begin{equation}
    \begin{aligned}
\tilde{h}(t)\simeq
&\frac{\omega_{x}}{4}(J_{1}(A)\sigma_{+}+J_{1}(A)\sigma_{-}),
\\
=&\omega_{x}J_{1}(A)\sigma_{x}/2.
    \end{aligned}
\end{equation}
The effective nutation frequency of a WOLF pulse is thus given by
\begin{equation}
    \begin{aligned}
\wSTnut=\omega_{x}J_{1}(A)=2\pi\sqrt{J^{2}_{13}+J^2_{23}}\sin(\phi+\theta/2)J_{1}(A).
    \end{aligned}
\end{equation}

\section{Technical Details}
\subsection{Samples}

The precursor solution for fumarate was prepared by dissolving 100~mM disodium acetylene dicarboxylate, 100~mM sodium sulfite, and 6~mM \RuCat (CAS number: 99604-67-8) in D$_2$O, heating to 60\degC, and passing through a Millex\textregistered\ 0.22~$\mu$m PES filter.

The precursor solution for maleate was prepared by dissolving 100~mM acetylene dicarboxylic acid and 5~mM \RhCat (CAS number: 7440-16-6) in methanol-d$_4$.  All materials and consumables were purchased from Merck.

Under the reaction and solvent conditions used for the experiments, maleate is expected to exist mainly in the form of the protonated singly-charged mono-hydrogen maleate anion, while fumarate is expected to exist as the doubly-charged non-protonated fumarate anion. These protonation states are ignored in the current report, for the sake of simplicity.

\subsection{Equipment}

\Para-enriched hydrogen was  produced by passing hydrogen gas over an iron oxide catalyst packed in 1/4 inch 316L stainless steel tubing cooled by liquid nitrogen.
The gas is bubbled through the solutions using a 1/16 inch PEEK capillary tube inserted inside a thin-walled Norell\textregistered\ pressure NMR tube. The Rheodyne MXP injection valves and the Keysight 33500B waveform generator were triggered and controlled by an Arduino Mega 2560 micro-controller board. The waveform generator was connected to the 3~cm wide and 30~cm long solenoid coil of 300 turns placed in the TwinLeaf~MS-4 mu-metal shield and used for generating the oscillating magnetic field.
The bias field was generated by the built-in Helmholtz coil of the Twinleaf shield, powered by a Keithley 6200 DC current source. 
A $\sim$200 turn solenoid guide coil was wound around the orifice penetrating the mu-metal shield and continuously driven by a second Keithley 6200 DC current source to produce $6~\mu$T field. The guide coil was used to avoid zero-field crossings during sample transportation.

\subsection{Experimental Procedure}

Before each experiment, 250~$\mu$L of sample was taken from a 8~mL stock solution and heated in a water bath to a temperature of 90\degC. 
The pressure tube was then quickly placed into the magnetic shield and the hydrogenation experiment was initiated.
The bubbling time $\tau_b$ was set to 10~s and 30~s for experiments involving maleate and fumarate, respectively. 
After 5 seconds during which polarisation transfer driven by oscillating field is performed, the sample was manually removed from the magnetic shield and inserted into the Oxford 400~MHz magnet equipped with a Bruker Avance Neo spectrometer.

The \Cth free-induction decays were excited by a hard pulse of 14.7~kHz rf amplitude and recorded with 65~k point density at the spectral width of ~200~ppm. Additional \Hone decoupling was used for all experiments. Thermal equilibrium \Cth spectra were recorded at room temperature with recycle delays of 120~s, averaging the signals from 512 and 360 transients for maleate and sodium fumarate, respectively.

\subsection{Frequency optimisation}

\begin{figure}[h]
\includegraphics[width=0.8\textwidth]{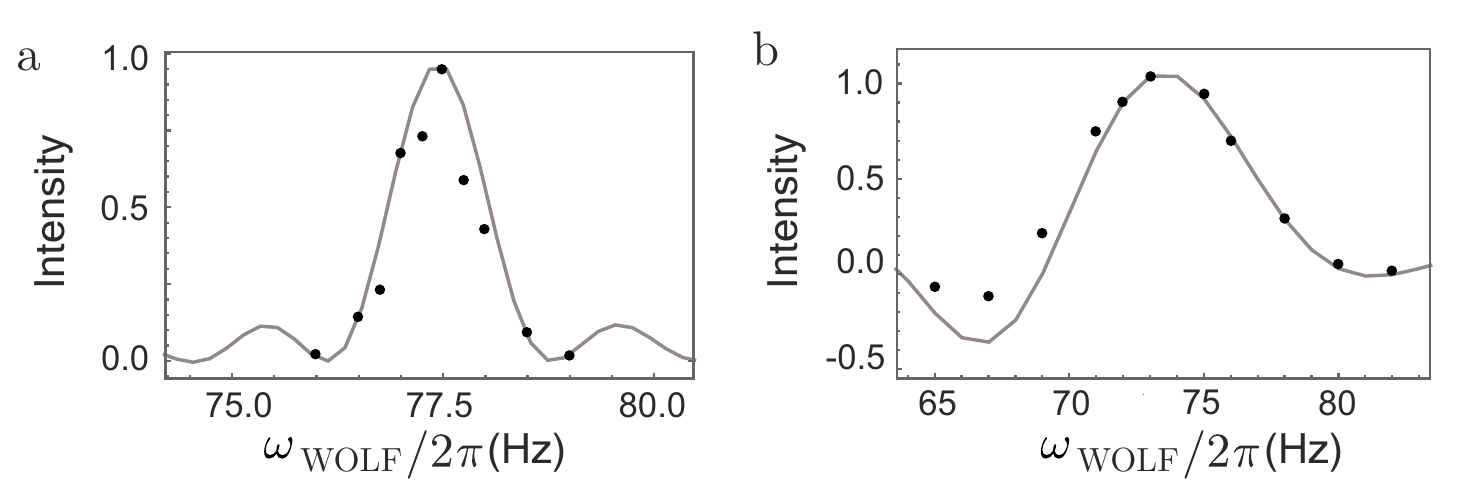}
\caption{\label{fig:frequency} Hyperpolarised fumarate (a) and maleate (b) \Cth signal intensity as a function of $\wWOLF$ at fixed duration $\tau$ equal to 50 and 9 magnetic field oscillations during the WOLF pulse for fumarate and maleate, respectively. Black dots depict experimental data whereas grey lines represent numerical simulations using the spin-spin coupling parameters found in the manuscript.
}
\end{figure}

As discussed in section \ref{seq:dynamics} and evident from equation \ref{eq:tilted_mat} the singlet-triplet transition is effectively driven once the condition $\wWOLF=\wST$ is fulfilled. The experimental optimisation of $\wWOLF$ is given in the Fig.~\ref{fig:frequency} which was performed at fixed WOLF pulse duration.
The experimental data points indicate well defined peaks with FWHM of $\sim 0.4$~Hz and $\sim 6$~Hz for fumarate and maleate, respectively.
The match between numerically simulated profiles and the data suggest that field inhomogeneity is negligible. 
The discrepancy in resonance width could be associated with the different number of magnetic field oscillations required to reach maximum signal at $\tau=\pi/(2\wSTnut)$.

